\documentclass[12pt,preprint]{aastex}
\newcommand{\msun}{{\rm \ M_\odot}}

\def\ie{{\it i.e.},\ } 
\def\vs{{\it vs.}\ } 
\def\eg{{\it e.g.},\ } 
\def\cf{{\it cf.}}

\begin{document}
\title{OGLE-2017-BLG-0373Lb: A Jovian Mass-Ratio Planet Exposes A New Accidental Microlensing Degeneracy}

\author{\textsc{
J. Skowron$^{1}$, 
Y.-H. Ryu$^{2}$, 
K.-H. Hwang$^{2}$, 
\and 
A. Udalski$^{1}$,
P. Mr\'{o}z$^{1}$, 
S. Koz{\l}owski$^{1}$,
I. Soszy\'{n}ski$^{1}$, 
P. Pietrukowicz$^{1}$, 
M. K. Szyma\'{n}ski$^{1}$, 
R. Poleski$^{1,3}$, 
K. Ulaczyk$^{1}$,
M. Pawlak$^{1}$,
K. Rybicki$^{1}$,
P. Iwanek$^{1}$\\
(OGLE Collaboration)\\
M. D. Albrow$^{4}$,
S.-J. Chung$^{2,5}$, 
A. Gould$^{2,3,6}$, 
C. Han$^{7}$, 
Y. K. Jung$^{8}$,
I.-G. Shin$^{8}$, 
Y.~Shvartzvald$^{9,^{\dag}}$, 
J. C. Yee$^{8}$, 
W.~Zang$^{10,11}$, 
W. Zhu$^{12}$, 
S.-M. Cha$^{2,13}$, 
D.-J. Kim$^{2}$, 
H.-W. Kim$^{2}$, 
S.-L. Kim$^{2,5}$, 
C.-U. Lee$^{2,5}$,
D.-J. Lee$^{2}$, 
Y. Lee$^{2,13}$, 
B.-G. Park$^{2,5}$,
R. W. Pogge$^{3}$ \\
(KMTNet Collaboration)\\
} 
}

%----------------------------------------------------------------
\affil{$^{1}$Warsaw University Observatory, Al. Ujazdowskie 4,
00-478 Warszawa, Poland}

\affil{$^{2}$Korea Astronomy and Space Science Institute, Daejon
34055, Korea}

\affil{$^{3}$Department of Astronomy, Ohio State University, 140 W.
18th Ave., Columbus, OH 43210, USA}

\affil{$^{4}$University of Canterbury, Department of Physics and
Astronomy, Private Bag 4800, Christchurch 8020, New Zealand}

\affil{$^{5}$Korea University of Science and Technology, Daejeon 34113, Korea}

\affil{$^{6}$Max-Planck-Institute for Astronomy, K\"{o}nigstuhl 17,
69117 Heidelberg, Germany}

\affil{$^{7}$Department of Physics, Chungbuk National University,
Cheongju 28644, Republic of Korea}

\affil{$^{8}$ Harvard-Smithsonian CfA, 60 Garden St., Cambridge, MA 02138, USA}

\affil{$^{9}$Jet Propulsion Laboratory, California Institute of
Technology, 4800 Oak Grove Drive, Pasadena, CA 91109, USA}

\affil{$^{10}$Physics Department and Tsinghua Centre for
Astrophysics, Tsinghua University, Beijing 100084, China}

\affil{$^{11}$Department of Physics, Zhejiang University, Hangzhou,
310058, China}

\affil{$^{12}$Canadian Institute for Theoretical Astrophysics, 
University of Toronto, 60 St George Street, Toronto, ON M5S 3H8, Canada}

\affil{$^{13}$School of Space Research, Kyung Hee University,
Yongin, Kyeonggi 17104, Korea}

\affil{$^{\dag}$NASA Postdoctoral Program Fellow}

\begin{abstract}
We report the discovery of microlensing planet
  OGLE-2017-BLG-0373Lb. We show that while the planet-host system has
  an unambiguous microlens topology, there are two geometries within
  this topology that fit the data equally well, which leads to a
  factor 2.5 difference in planet-host mass ratio, \ie $q=1.5\times
  10^{-3}$ \vs $q=0.6\times 10^{-3}$. We show that this is an
  ``accidental degeneracy'' in the sense that it is due to a gap in
  the data. We dub it ``the caustic-chirality degeneracy''.  We trace
  the mathematical origins of this degeneracy, which should enable
  similar degenerate solutions to be easily located in the future. A
  Bayesian estimate, based on a Galactic model, yields a host mass
  $M=0.25^{+0.30}_{-0.15}\,M_\odot$ at a distance
  $D_L=5.9^{+1.3}_{-1.95}$~kpc. The lens-source relative proper motion
  is relatively fast, $\mu=9$~mas/yr, which implies that the host
  mass and distance can be determined by high-resolution imaging after
  about 10 years. The same observations could in principle resolve the
  discrete degeneracy in $q$, but this will be more
  challenging.
\end{abstract}


\section{Introduction\label{sec:intro}}
The interpretation of gravitational microlensing events is subject to
a variety of discrete degeneracies, which can be divided into two
broad categories: intrinsic and accidental. Intrinsic degeneracies
are generally rooted in symmetries of the lens equation.  The most
famous of these is the ``close/wide'' degeneracy between binary lens
solutions having projected separations $s$ (scaled to the angular
Einstein radius $\theta_{\rm E}$) that approximately obey $s\leftrightarrow
s^{-1}$. These were first discovered in the context of a theoretical
study of planetary lensing in the high-magnification regime by
\citet{griest98}.  Soon after, \citet{dominik99} showed that they
constituted the limiting form of a degeneracy that impacted binary
lenses of all mass ratios and was rooted in a symmetry between a
quadrupole expansion of the lens equation for $s<1$ and a shear
expansion for $s>1$.  \citet{an05} then found that this degeneracy
continued at next order when he studied the lens equation still more
deeply.

Another intrinsic degeneracy is the so-called ``ecliptic degeneracy''.
Its origins were first studied mathematically by \citet{smp03}, whose
work was later extended in different directions by \citet{gould04} to
the ``jerk-parallax degeneracy'' and by \citet{ob09020} to a discrete
degeneracy involving four parameters in binary lensing.

On the other hand, there are also accidental degeneracies, \ie those
that do not derive from deep symmetries of the lens equation but
rather from ``accidental'' alignments of magnification patterns
arising from unrelated lens configurations. Very often, these
magnification-pattern alignments are restricted to a limited domain,
so that the degeneracy is easily broken if there are sufficient data
in other regions of the light curve where this alignment
fails. However, there are at least two degeneracies that can be quite
intractable even though they are not rooted in the lens equation. One
of these, discovered by \citet{gaudi98}, is between planetary events (two
lenses one source, 2L1S) that lack detailed caustic crossings and
binary-source events (1L2S). An especially complex example of this was
discovered by \citet{ob151459}, who found equally good 3L1S, 2L2S,
and 1L3S solutions for OGLE-2015-BLG-1459. In the end, this degeneracy
was resolved in favor of the triple source (1L3S) solution, but only
because there were color data serendipitously available, and the 1L3S
solution predicted strong chromatic effects while the others did not.

Another quite intractable (and also quite unexpected) degeneracy was
recently found by \citet{ob151459} for OGLE-2017-BLG-0173 between
two classes of ``Hollywood'' events in which the source is of order or
larger than the planetary caustic \citep{gould97}. They showed that the
``Cannae'' and ``von Schlieffen'' topologies\footnote{The terms
  ``Cannae'' and ``von Schlieffen'' coined by \citet{ob170173} come
  from two distinct military tactics. The first is the
  double-envelopment tactic (strong flanks and weak center) used by
  Hannibal in the Battle of Cannae between the armies of Carthage and
  the Roman Republic, 2 August 216 BCE in Apulia. The second is named
  after the single-envelopment (strong right flank and weak left
  flank) offensive plan devised by Field Marshal Alfred von Schlieffen
  (the Chief of the Imperial Army German General Staff, 1891--1906),
  which variation was used in the German invasion of France and
  Belgium in August 1914.}, in which the source fully or partially
envelops the caustic, produced almost identical light curves. In
fact, this degeneracy is not intrinsically intractable: the two models
predict different light curves in regions where there are no data.
Still, these predicted differences are remarkably small.  The nature
of this degeneracy is still not fully understood. \citet{ob170173}
found that it did not apply to another Cannae Hollywood event
OGLE-2005-BLG-390 \citep{ob05390}, while \citet{ob171434}
found that it did apply to hypothetical events similar to
OGLE-2017-BLG-0173 but with much lower mass ratio.

It is exceptionally important to identify as many classes of these
degeneracies as possible so that the true solution is not missed and
so that the cases for which there are multiple viable solutions with
substantially different physical implications are recognized.  For
example, prior to the discovery of the jerk-parallax degeneracy,
\citet{alcock01} found a microlens parallax solution for
MACHO-LMC-5 that led to the conclusion of a luminous lens at a
distance of 200~pc but with a mass well below the hydrogen-burning
limit, which they regarded as quite remarkable. The second parallax
solution was found by \citet{gould04} accidentally, in the course of
trying to reproduce this solution, and he proceeded to identify its
mathematical nature within the framework of the \citet{smp03}
formalism. Part of that work yielded an equation (made substantially
more user-friendly by \citealt{mb03037}) to predict the jerk-parallax
counterpart once one solution is found. Hence, at this point, one can
easily determine whether this degeneracy is present.

Similarly, a downhill approach could easily miss (and, actually, did
originally miss) the Cannae solution to OGLE-2017-BLG-0173, being
``blocked'' by the ``ne\-arby'' von Schlieffen solution. In fact, while
these solutions are ``nearby'' in most lensing parameters, they differ
by a factor $\approx2.5$ in planet-host mass ratio. Indeed, \citet{ob170173}
showed analytically that von Schlieffen mass ratios
should generically be higher by a factor $\approx2$. Hence, in the
future, these degeneracies can be easily detected because if either
the Cannae or von Schlieffen solutions are found, the parameters of
the other can be approximately predicted.

Here we analyze OGLE-2017-BLG-0373 and show that it exhibits a
previously unknown accidental degeneracy, which we dub ``the
caustic-chirality degeneracy''. We investigate the origins of this
degeneracy and show that it can easily appear in events for which half
or more of the light curve traversing a planetary caustic is devoid of
observational data.

\section{Observations\label{sec:obs}}

OGLE-2017-BLG-0373 is at ${\rm (RA,Dec)}=(17{:}57{:}19.06,-31{:}57{:}06.2$),
corresponding to $(l,b)=(-1.31,-3.71)$. It was discovered and
announced as a probable microlensing event by the OGLE Early Warning
System \citep{ews1,ews2} at UT 15:04 23 March
2017\footnote{\it http://ogle.astrouw.edu.pl/ogle4/ews/2017/blg-0373.html}.
The event lies in OGLE-IV field BLG507 \citep{ogleiv}, for which
OGLE observations were at a cadence of $\Gamma=0.4~{\rm hr^{-1}}$
using their 1.3~m telescope at Las Campanas, Chile.

The Korea Microlensing Telescope Network (KMTNet, \citealt{kmtnet})
observed this field from its three 1.6~m telescopes at CTIO (Chile,
KMTC), SAAO (South Africa, KMTS) and SSO (Australia, KMTA), in its two
slightly offset fields BLG01 and BLG41, with a combined cadence of
$\Gamma=4~{\rm hr^{-1}}$. However, the KMTA data are of substantially
lower quality, and they do not contribute to characterizing the
planetary anomaly. Hence, we do not use them. The event was identified
as SAO41N0303.005729 by KMTNet.

The great majority of observations were carried out in the {\it
I}-band with occasional {\it V}-band observations made solely to
determine source colors. All reductions for the light curve analysis
were conducted using variants of difference image analysis (DIA,
\citealt{alard98}), specifically \citet{wozniak2000} and \citet{albrow09}.

For the modeling we take 100 days-long light curves from KMTNet
observatories covering the whole event, and 1000 days-long OGLE-IV
light curve (2015-2017), and use it as a flux reference.

After fitting a simple microlensing model, we can rescale the
error-bars in order to achieve $\chi^2$ per degree of freedom equal
to~1. This is warranted for three reasons. First, we are confident
about the microlensing nature of the event. Second, we see increased
scatter of residuals compared to the instrumental error-bars on all
parts of the light curve, even smooth and uneventful. And third, we
expect underestimated errors based on many other microlensing and also
unrelated studies (\eg \citealt{skowron16}). Moreover, this procedure
is needed as it helps making estimates of uncertainties of derived
parameters more robust. For error-bar rescaling we use the standard
formula: $\sigma=k(\sigma_0^2+\sigma_{\rm min}^2)^{1/2}$. The scalling
factor $k$ is set to be 1.3483, 1.9707, 1.8887, 2.0110, 2.0875 for
OGLE, KMTC(BLG01), KMTC(BLG41), KMTS(BLG01), and KMTS(BLG41) datasets,
respectively, while $\sigma_{\rm min}$ is set to 1~mmag.

The light curves used in this work will be available from NASA
Exoplanet Archive ({\it https://exoplanetarchive.ipac.caltech.edu/}) and
from the OGLE website ({\it http://ogle.astrouw.edu.pl/}).

\section{Light Curve Analysis\label{sec:anal}}

\subsection{Heuristic Analysis\label{sec:heuristic}}

\begin{figure}
\plotone{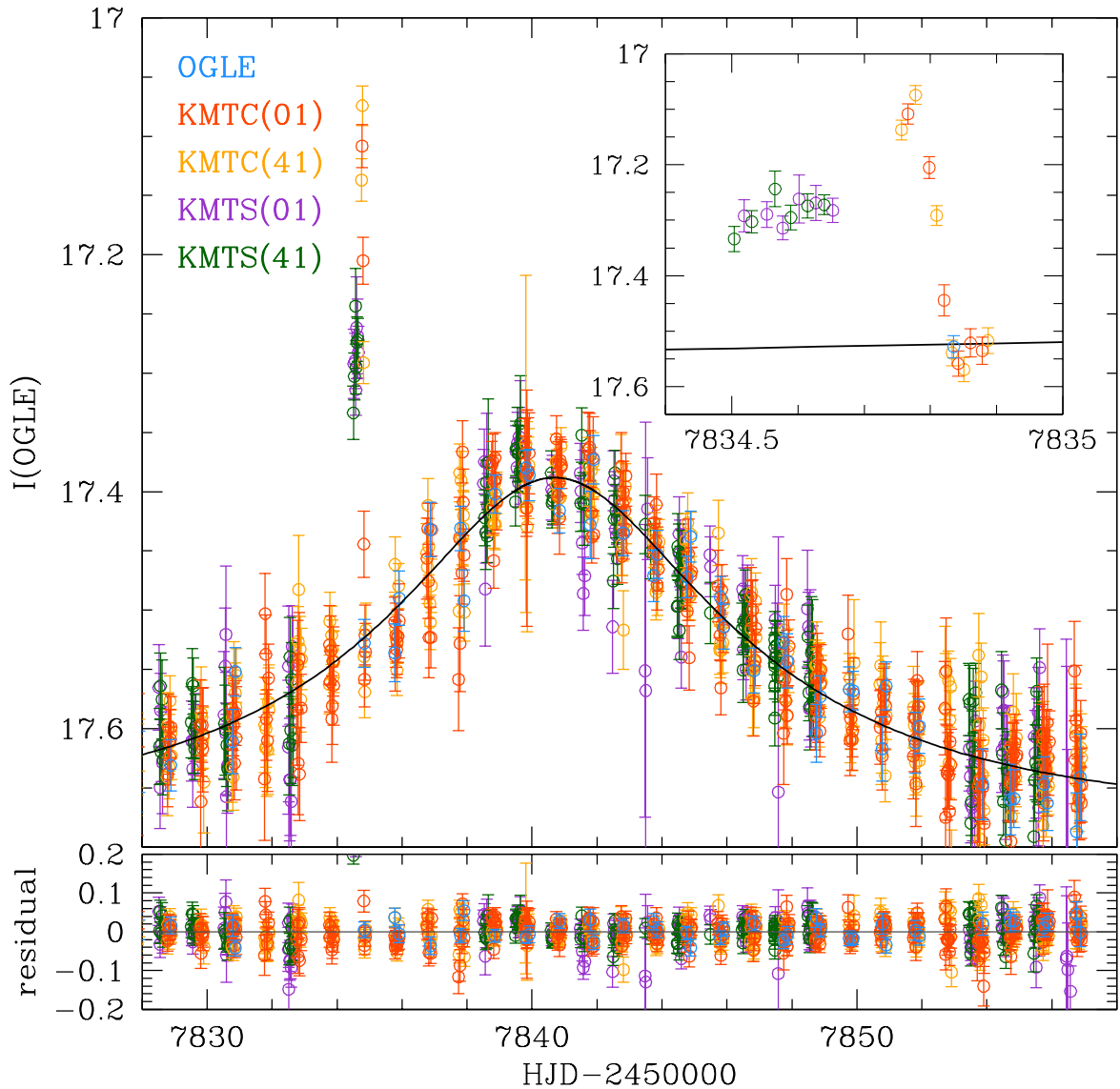}
\caption{Light curve and point-lens \citet{pac86} model for
  OGLE-2017-BLG-0373. The data points are color-coded by observatory
  and field. The inset shows the anomaly. Even without a model, it is
  clear that this is a caustic trough followed by a caustic exit.
  While this morphology is eerily reminiscent of (a time reversal of)
  \citet{gouldloeb} original illustration of planetary-caustic
  microlensing, it is actually broadly consistent with five different
  binary-microlensing topologies, which must be systematically
  vetted. See Fig.~2.}
\end{figure}

Fig.~1 shows the data, together with a point-lens model, for
OGLE-2017-BLG-0373.  With the exception of an anomaly lasting a total
of just 0.3~d near ${\rm HJD}^\prime\equiv{\rm HJD}-2450000=7835$,
the event appears to be a simple point-lens \citet{pac86} light
curve, characterized by three parameters: the time of closest approach
$t_0$, the impact parameter (normalized to $\theta_{\rm E}$) $u_0$, and the
Einstein timescale $t_{\rm E}$. The apparent amplitude is only $\Delta
I\approx0.3$~mag, but a priori one does not know whether this is
because the event has an intrinsically low amplitude or is heavily
blended.  We exclude the anomaly and fit to a point lens, thereby
finding $(t_0,t_{\rm eff},t_{\rm E})=\break[(7840.70,4.73,12.9)\pm 
(0.05,0.22,1.2)]$~d where $t_{\rm eff}\equiv u_0\,t_{\rm E}$.

\begin{figure}
\plotone{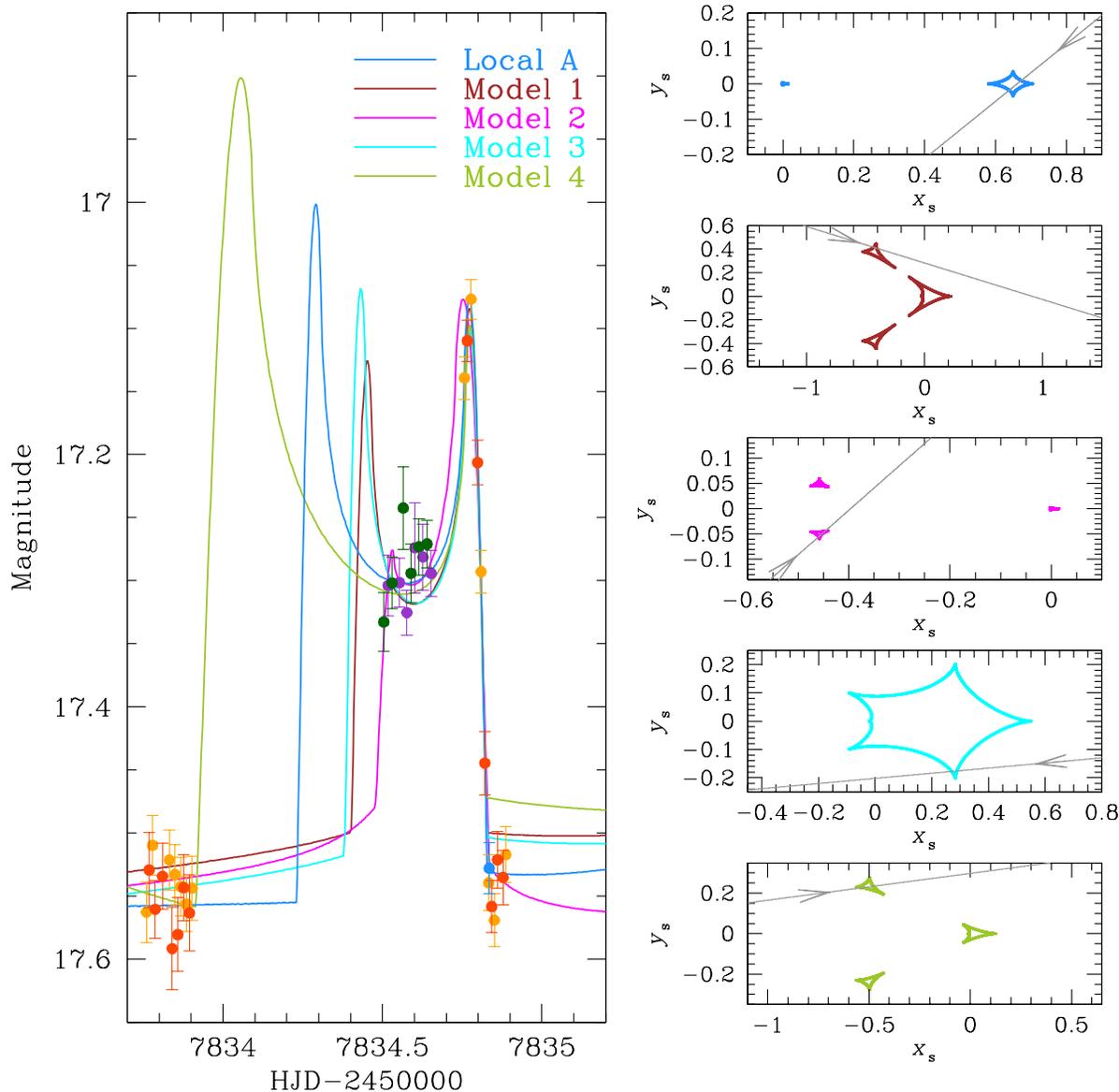}
\caption{Models derived from five different binary-lens topologies that
  roughly fit the light curve of OGLE-2017-BLG-0373. All look roughly
  plausible, but one of them (the major-image planetary-caustic model
  (Local A) is favored over all the others by
  $\Delta\chi^2\gtrsim100$. See Table~1. The caustic geometries for
  the models are shown in the five {\it right-hand panels}, with the
  caustics color-coded to match the models in the {\it left-hand panel}.
  Note that all of the caustic diagrams are drawn to the same scale.}
\end{figure}

The entire light curve, with its clear caustic exit superposed on a
\citet{pac86} curve, is remarkably suggestive of Fig.~1 from
\citet{gouldloeb}, \ie the original illustration of a
planetary-caustic perturbation to a microlensing event.  Because of
this classic appearance, the event was immediately singled out for
further analysis when it was noticed in the course of testing a
``quick look'' KMTNet pipeline to aid in the choice of Spitzer
microlensing targets \citep{prop2016}\footnote{The event could not
itself be targeted for Spitzer observations because it was already
at baseline at the onset of the Spitzer campaign.}.

However, in contrast to the \citet{gouldloeb} figure (which was
purely theoretical), the OGLE-2017-BLG-0373 light curve is ``missing''
its caustic entrance. For this reason, the light curve is actually
roughly consistent with a remarkably wide range of binary-lens
topologies. See Fig.~2. A more detailed analysis shows that all but
one of these five basic topologies is ruled out at the
$\Delta\chi^2\gtrsim100$ level. Nevertheless, it is important to note
that if the actual event had been generated by any of the other models
shown in Fig.~2, it would have appeared qualitatively the same to the
eye, and so would have triggered the same investigation. Hence, it is
crucial to explore all possible topologies, even when an event looks
``obviously planetary''.

We next undertake a heuristic analysis of the surviving topology, \ie
a planetary perturbation of the major image $(s>1)$, which is in fact
the same as that illustrated by \citet{gouldloeb}. This analysis
will frame the introduction of a new degeneracy within this topology.

The anomaly is offset from the peak by $\delta t=-6.2$~d. Hence,
the angle of the source trajectory relative to the binary axis is
$\alpha=\tan^{-1}(t_{\rm eff}/\delta t)=143^\circ\pm2^\circ$. The
lens-source separation (normalized to $\theta_{\rm E}$) at the time of
the anomaly is $u_{\rm caust}=\sqrt{t_{\rm eff}^2+(\delta t)^2}/t_{\rm
E}=0.605\pm0.055$.  Noting that $s-s^{-1} =u_{\rm caust}$ for
major-image perturbations, we obtain $s=1.35\pm0.03$.

To estimate the planet-host mass ratio $q$, we apply the formalism of
\citet{han06}, specifically Section~3.1. He gives the caustic major and
minor axes as $(\Delta\xi_c,\Delta\eta_c)=\sqrt{16q/(s^4\mp s^2)}
\rightarrow q^{1/2}(3.27,1.76)$. Unfortunately, we do not have a very
good estimate of the duration of the time that the source is inside
the caustic, $\Delta t_{\rm caust}$, because there are no data from
the caustic entrance. Nevertheless, there is a clear upper limit from
the flat KMTC data at HJD$^\prime=7834.90$, and a plausible lower
limit from the fact that the KMTS data are roughly flat
HJD$^\prime=7834.45$. Hence we estimate, $0.6~{\rm d}<\Delta t_{\rm
caust}<0.9~{\rm d}$. The main difficulty in estimating $q$,
however, is that it is hard to assess the geometry of the crossing
given the fragmentary light curve. We do know that the angle is
$\alpha\simeq143^\circ$, which implies that the caustic is being
crossed diagonally. Hence, the length of the transit will be
substantially shorter than the major axis. We adopt the minor axis.
We then obtain $q=[(\Delta t_{\rm caust}/t_{\rm E})/\Delta\xi_c]^2
\simeq1\times 10^{-3}$. Given the crude estimates of $\Delta t_{\rm
caust}$ and of the geometry of the crossing, we consider that this
should be accurate to a factor $\approx 3$. However, this exercise at
least tells us that this is a Jovian mass-ratio planet.

Finally, we estimate the source self-crossing time $t_*=\rho t_{\rm
E}$, where $\rho\equiv\theta_*/\theta_{\rm E}$ and $\theta_*$ is the
angular radius of the source. From Fig.~1, the time from the peak to
the end of the caustic exit is $\Delta t_{\rm exit}=0.073$~d. From
Fig.~1 of \citet{gouldandronov99} this is related to self-crossing
time by $t_*=\Delta t_{\rm exit}\sin\phi/1.7=1.0\sin\phi$~hr, where
$\phi$ is the angle between the source trajectory and the caustic.
Because the source crosses the caustic diagonally, we can estimate
$\sin\phi\approx1$. Hence, $t_*\simeq1$~hr and $\rho\simeq0.003$.

\subsection{Numerical Analysis\label{sec:numerical}}

\begin{deluxetable}{lccc}
\tablecolumns{4}
\tablewidth{0pc}
\tablecaption{\textsc{Five Candidate Models}}
\tablehead{ \colhead{Model } & \colhead{$s$} & \colhead{$q$} & \colhead{$\chi^2$} } 
\startdata
  Local A     &1.38        &0.0015       &2795.0  \\
  Model 1     &0.79        &0.056        &2890.9  \\
  Model 2     &0.80        &0.0009       &2893.8  \\
  Model 3     &1.15        &0.033        &2944.8  \\
  Model 4     &0.77        &0.019        &2967.4  \\
\enddata
\label{tab:5mod}
\end{deluxetable}

We search for solutions in two stages.  First we conduct a grid search
over $(s,q)$ space, in which these two parameters are held fixed,
while the remaining five parameters $(t_0,u_0,t_{\rm E},\rho,\alpha)$
are allowed to vary in a Monte Carlo Markov Chain (MCMC). The first
four of these are seeded at the values derived in Section~3.1, while
the last is seeded on a grid of six values. To evaluate the
magnification, we use inverse ray shooting in and near the caustic
\citep{kayser86,schneider88,wambs97} and
multipole approximations \citep{pejcha09,gould08}
elsewhere. In the second step, we seed each MCMC run with the
parameters of the local minima identified in the first step. Fig.~2
shows the models from these refined minima. As discussed in
Section~3.1, there are five different models, each with very different
topologies, that yield plausible likenesses to the data.  However, one
of these is favored by $\Delta\chi^2\gtrsim100$ relative to the
others, namely the major-image planetary-caustic model, which
corresponds to the lens-system topology presented in the top right
panel of Fig.~2. In Table~1, we list the $(s,q)$ and $\chi^2$ of
these five models.

Where do Models 1--4 fail?  Fig.~3 displays the residuals to the five
models shown in Fig.~2. All four failed topologies suffer from somewhat
``wavy'' residuals over many weeks. In the case of Model~2 and Model~4,
this is punctuated by a severe discrepancy on the day following the caustic
exit (and the day of the caustic exit for Model~4).  However, Model~1 and
Model~3 fail due to roughly one week periods of systematic residuals,
centered on ${\rm HJD}^\prime\approx7834$ and $\approx7823$,
respectively.

\begin{figure}
\plotone{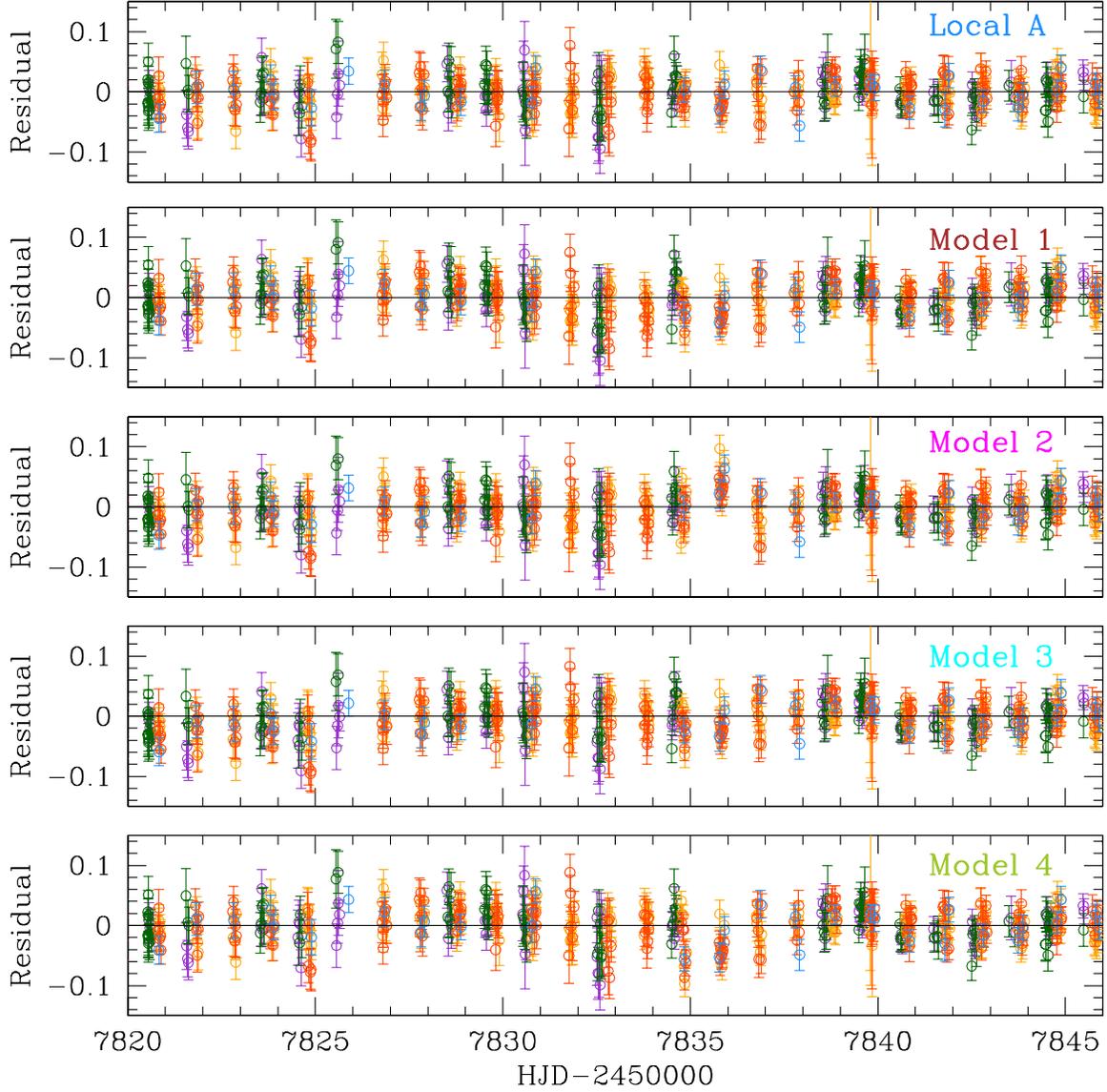}
\caption{Residuals for the five models shown in Fig.~2.
Model~1 and Model 3 fail primarily because of week-long systematic
residuals centered on ${\rm HJD}^\prime\approx7834$ and $\approx7823$,
respectively. Models~2 and 4 fail primarily due to concentrated
deviations on and/or immediately after the caustic exit.
The peak of the event is $t_0\approx7840.6$ and anomaly is
at $t\approx7834.5$.}
\end{figure}

\begin{deluxetable}{lcc}
\tablecolumns{3}
\tablewidth{0pc}
\tablecaption{\textsc{Best-fit solutions}}
\tablehead{ \colhead{Parameters } & \colhead{Local A} & \colhead{Local B}} 
\startdata
  $\chi^2/\rm{dof}$               &2795.00/2795        &2795.22/2795        \\
  $t_0$ $(\rm{HJD}^{\prime})$     &7840.607 $\pm$ 0.054 &7840.636 $\pm$ 0.054 \\
  $u_0$                           &0.416 $\pm$ 0.011    &0.371 $\pm$ 0.015    \\
  $t_{\rm E}$ $(\rm{days})$       &11.993 $\pm$ 0.243   &12.849 $\pm$ 0.349   \\
  $s$                             &1.376 $\pm$ 0.010    &1.349 $\pm$ 0.011    \\
  $q$  $(10^{-3})$                &1.544 $\pm$ 0.352   &0.651 $\pm$ 0.113   \\
  $\alpha$ $(\rm{rad})$           &2.461 $\pm$ 0.015    &2.465 $\pm$ 0.011    \\
  $\rho$ $(10^{-3})$              &2.935 $\pm$ 0.190    &2.466 $\pm$ 0.201    \\
  $f_S$                           &0.263 $\pm$ 0.009    &0.224 $\pm$ 0.012    \\
  $f_B$                           &1.086 $\pm$ 0.009    &1.125 $\pm$ 0.012    \\
    \cline{1-3}
  $t_*$ $(\rm{days})$             &0.035 $\pm$ 0.002    &0.032 $\pm$ 0.002   \\
  $t_{\rm eff}$ $(\rm{days})$     &4.988 $\pm$ 0.089    &4.765 $\pm$ 0.100   \\
\enddata
\label{tab:ulens}
\tablecomments{Note: The fluxes $f_S$ and $f_B$ are normalized to $I=18$~mag in OGLE-IV instrumental magnitude scale, or $I_{\rm std}=18.114$~mag in calibrated {\it I}-band magnitudes.}
\end{deluxetable}

In the above fitting process, the preferred topology has a well-localized
minimum, which is given as ``Local A'' in Table~2.  Note that the heuristic
analysis of Section~3.1 quite accurately predicted all the parameters
except $q$. This shows that the overall geometry (for the major-image
planetary-caustic topology) is very well constrained by the gross features
of the light curve. However, it leaves open the question of whether the
specific value of $q$ is actually correct.

Indeed, when we run a hotter chain (\ie with error bars artificially
inflated), we find that there are actually two neighboring minima. These
have nearly identical $\chi^2$ but differ in mass ratio $q$ by a factor 2.5
(0.4 in the log).  Fig.~4 shows the models and caustic geometries of these
two solutions. Table~2 gives the parameters for both models.

\begin{figure}
\plotone{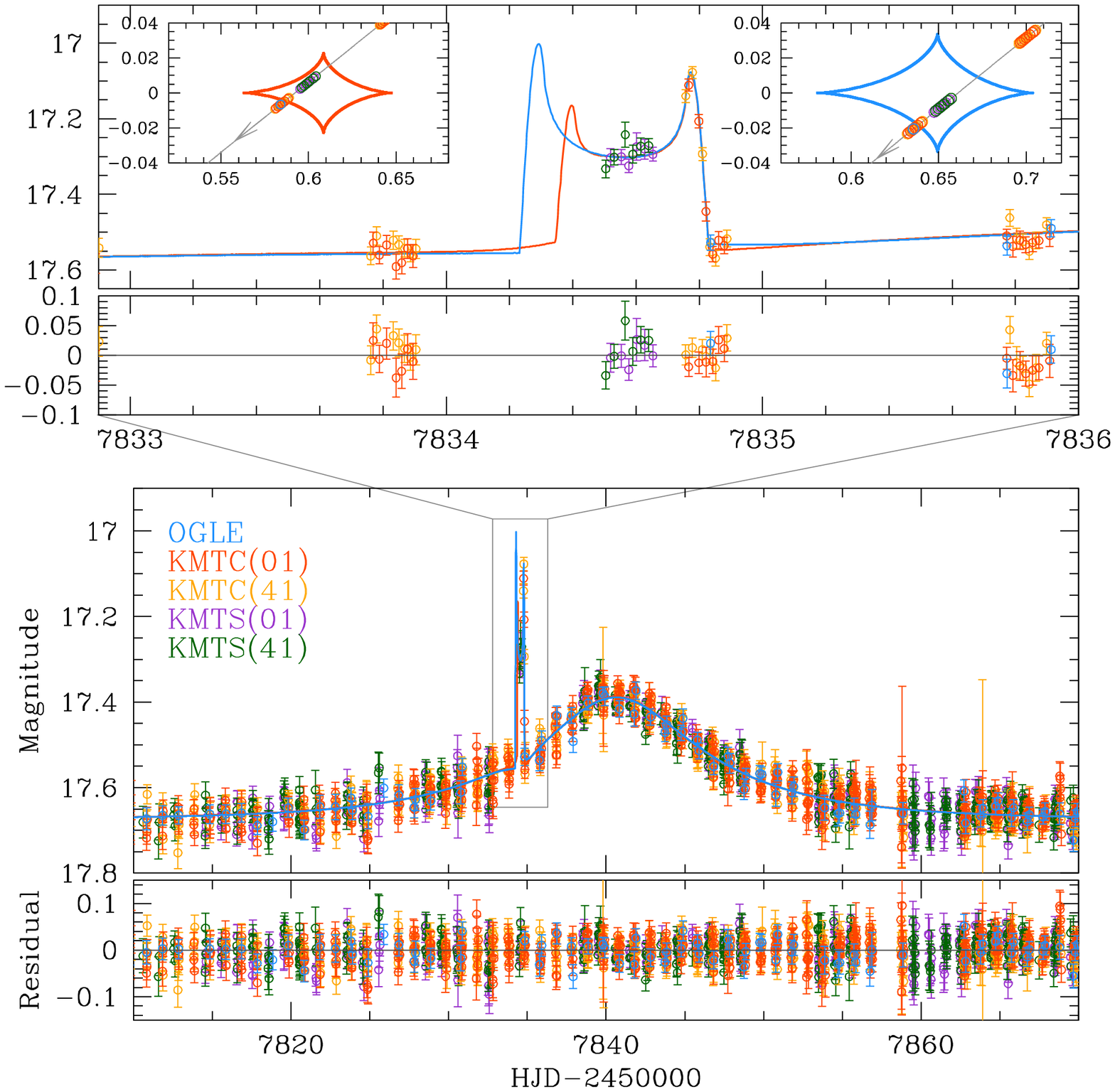}
\caption{Two models compared to the data for OGLE-2017-BLG-0373.
The two models, which have extremely similar parameters except for a
factor 2.5 difference in mass-ratio $q$, fit the data equally well.
This degeneracy is accidental in the sense that additional data
over the caustic entrance would clearly have distinguished between
them.  The caustic diagrams for each model are shown in the
insets with the same color coding as the model light curves.}
\end{figure}

\subsection{Origin of the Degeneracy\label{sec:origin}}

Why are there two solutions, rather than either one or many?  At one level,
the reason for multiple solutions is obvious from inspection of Fig.~4:
there are no data during or shortly after the caustic entrance. Even a few
data points would have clearly distinguished between these models.
However, this still does not explain why there are two solutions rather
than many discrete solutions or a large continuous degeneracy.

To investigate this question, we employ the $\Delta\xi$ parameter
introduced by \citet{ob170173} to analyze the degeneracy in the case of
OGLE-2017-BLG-0173, the offset of the source trajectory from the planetary
caustic,
$$\Delta\xi \equiv u_0\,{\rm csc}\,\alpha-\xi_+(s),\qquad\xi_+(s)\equiv s-s^{-1}\eqno(1)$$
where $\xi_+(s)$ is the source position for which the major image coincides
with the planet.  As in that study, this parameterization will enable us to
closely study the very small changes in source trajectory that lead to big
changes in physical implications. Fig.~5 shows $\Delta\chi^2$ as a
function of position on the $(\log q,\Delta\xi)$ plane and can be directly
compared to Fig.~4 of \citet{ob170173}.  This figure confirms what is
already apparent by comparing the two caustic diagrams in Fig.~4, namely
that the source crosses the binary axis on opposite sides of the caustic
center for the two models, and that it passes on the side closer to the
host for the case of smaller $q$.  However, it also shows that there is a
$\Delta\chi^2\approx9$ ``barrier'' between the two solutions at about
$\Delta\xi=-0.0075$.

\begin{figure}
\plotone{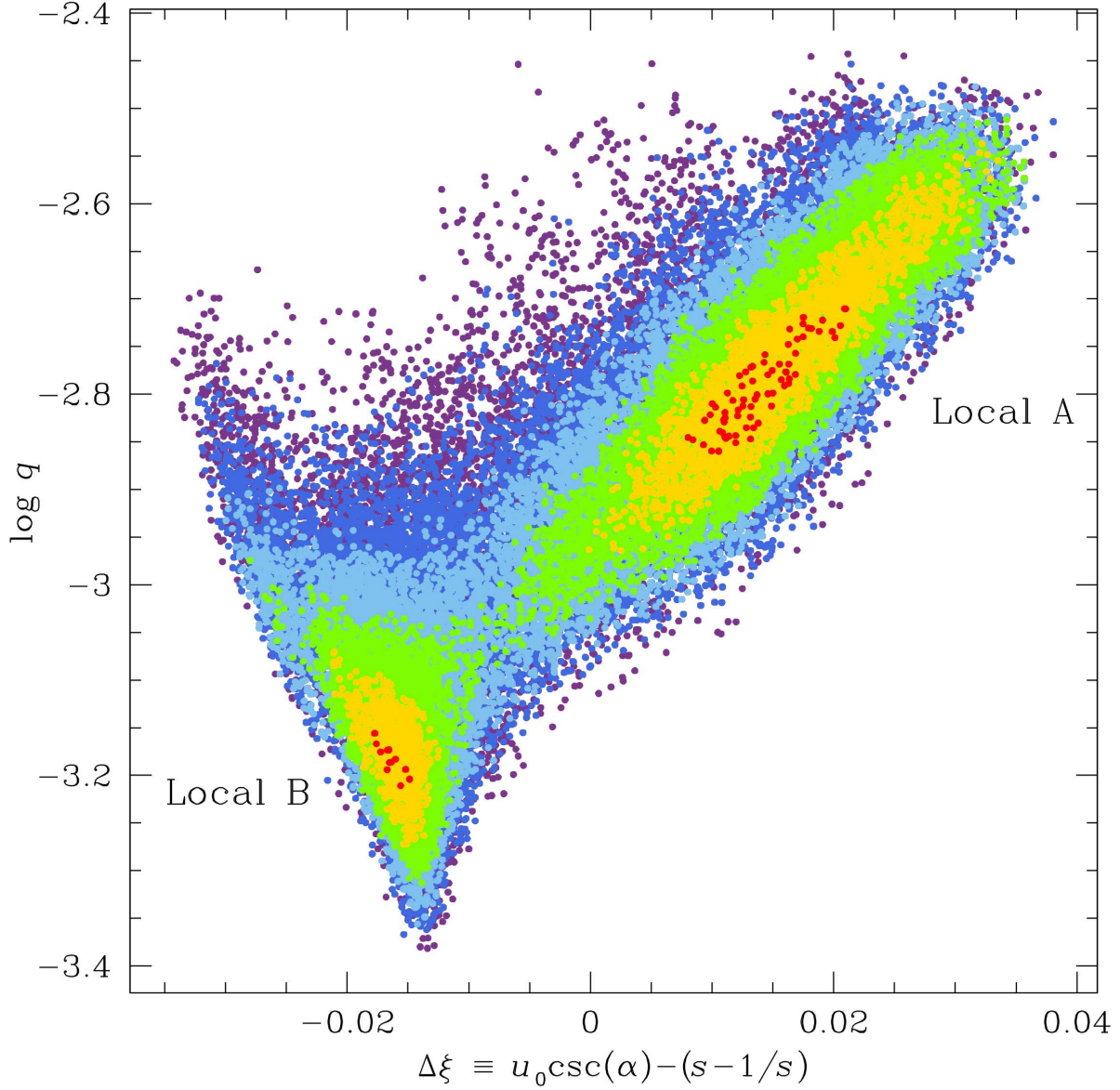}
\caption{$\chi^2$ distribution over the plane of $\log q$ \vs
  $\Delta\xi\equiv u_0\,{\rm csc},\alpha-(s-s^{-1})$, where $\Delta\xi$ is
  the offset between the source and the center of the caustic at the time
  that the source crosses the binary axis.  Red, yellow, green, cyan, blue,
  and purple represent $\Delta\chi^2<1,4,9,\ldots 36$. The two models
  have virtually identical $\chi^2$ but are separated by a barrier
  $\Delta\chi^2_{\rm barrier}\approx9$. The origin of this structure is
  investigated in Section~3.3.}
\end{figure}

Hence, we are led to ask why is the fit worse at this intermediate
position?  This is another way of asking why the degeneracy is two-fold?
Apart from the duration of the caustic exit, $\Delta t_{\rm exit}$, the
profile of a generic fold-caustic exit is basically set by three
parameters: the post-caustic magnification $A_0$, the time of the exit,
$t_{\rm 0,caust}$, and the ``strength'' of the caustic, $Z$, \ie
$A(t)=A_0+Z\;(t_{\rm 0,caust}-t)^{-1/2}\;\Theta(t_{\rm 0,caust}-t)$, where
$Z$ is constant and $\Theta$ is the Heaviside step function (\cf Eq.(8) of
\citealt{gouldandronov99}).  The functional form of the caustic-exit data (in
this case from KMTC) directly determine $A_0$, $t_{\rm 0,caust}$, and
$\Delta t_{\rm exit}$, which leaves only one degree of freedom: $Z$. The
caustic strength $Z$ generically peaks at the cusps and reaches a minimum
somewhere between them along the fold caustic.  Thus, for fixed $q$ there
are exactly two (or zero) positions along the fold caustic at which the
predicted peak of the caustic exit matches the observed peak.

Of course, if one has chosen the correct $q$ then it will be possible to
find the other correct parameters so the magnification evolution everywhere
correctly predicts the data. As the model moves away from this $q$, it will
still be possible to find the remaining parameters so that the caustic exit
is correctly predicted, but the resulting position of the caustic exit will
create increasing tension with the profile in the interior of the
caustic. In the present case, because of gaps in the data, particularly at
the caustic entrance, the tension is entirely generated by mis-predicting
the flux seen from KMTS near ${\rm HJD}^\prime=7834.5$. That is, as $q$
declines, the trajectory is driven closer to the cusp as it exits the fold
caustic, which implies that it is also traveling closer to the
(neighboring) fold caustic when it is interior to the caustic. This causes
the predicted flux in the interior to be too high.

Initially, at the correct $q$, the alternate allowed point of caustic exit
predicts a flux in the interior that is either too high or too low by some
arbitrary amount.  If it is too high, then as $q$ is adjusted upward, the
point of exit will be driven away from the cusp (by the overall increase of
strength $Z$ at higher $q$) and this will move the trajectory in the
interior away from the (neighboring) fold caustic, so that lower flux will
be predicted.  Eventually, at sufficiently high $q$, the prediction will
match the observations.

Hence, if there are two independent pieces of information from the
remainder of the caustic, then this discrete degeneracy will almost always
be broken (except for those due to deep symmetries, like $s\leftrightarrow
s^{-1}$). For example, a caustic entrance would have four independent
pieces of information. Inspection of Fig.~4 shows that even one of these
(such as the height or time of the entrance) would resolve the degeneracy.
However, with only one piece of information (as in the present case), the
continuous degeneracies are broken but a discrete degeneracy is very
likely.

From Fig.~5, we see that the two minima have values of $\Delta\xi$ with
opposite signs, meaning that the trajectories pass on opposite sides of the
nominal center of the planetary caustic.  That is, the trajectories have
opposite chirality with respect to the caustic center. This feature is
basically a product of the fact that the caustic exits must be on opposite
sides of the minimum strength of the caustic wall. Hence, it is likely to
be a generic (though probably not universal) feature of this degeneracy.
For this reason, we dub this accidental degeneracy as the
``caustic-chirality degeneracy''.

While Fig.~5 most directly illustrates the physical origins of the
caustic-chirality degeneracy (\ie a degeneracy between trajectories
passing on opposite sides of the caustic center), it is also of interest to
know how this degeneracy is projected onto the standard $(\log q,\log s)$
parameterization. We show this in Fig.~6.

\begin{figure}
\plotone{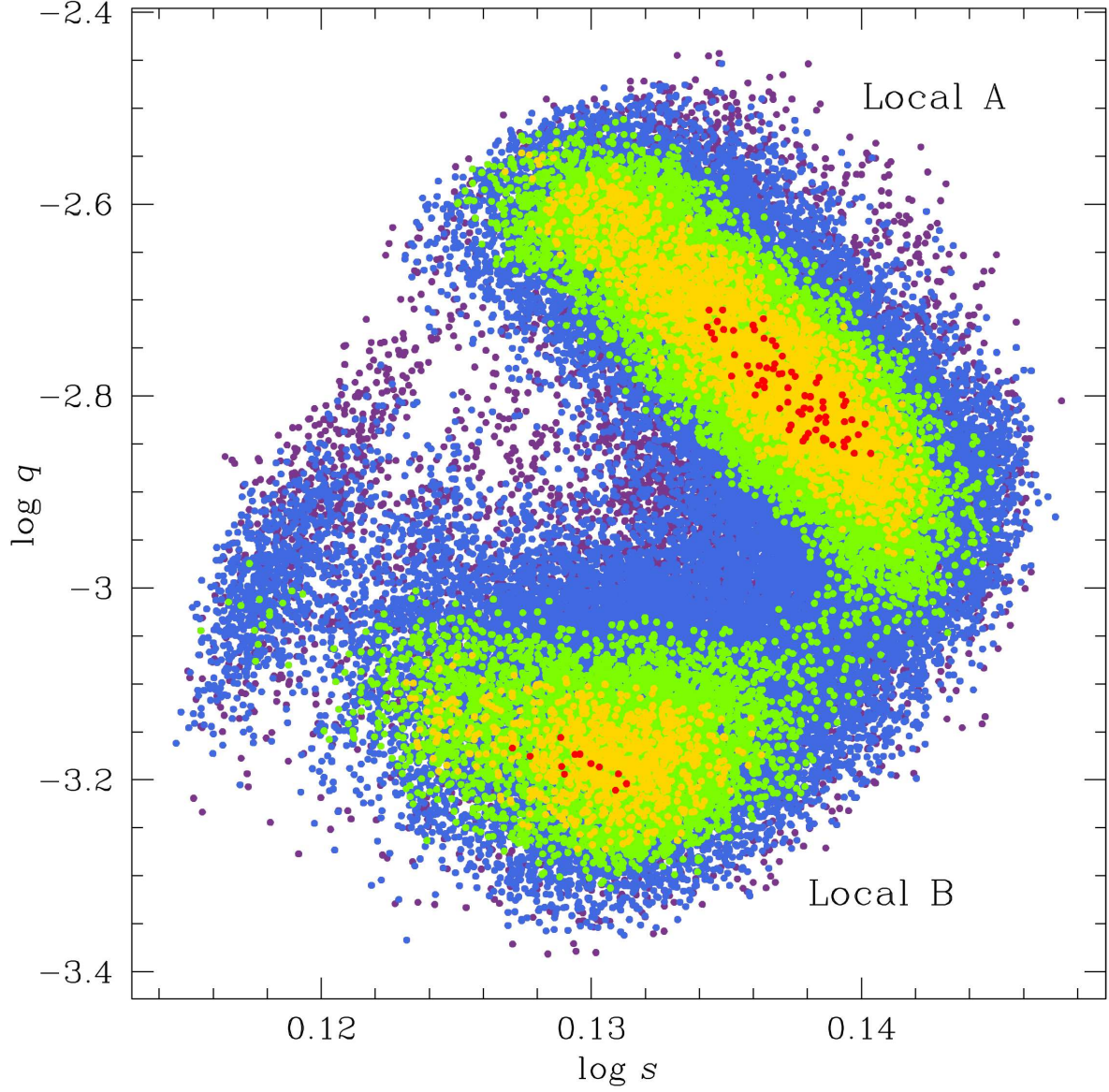}
\caption{$\chi^2$ distribution over the plane of $\log q$ \vs $\log s$.
As in Fig.~5, the red, yellow, green, cyan, blue, and purple represent 
$\Delta\chi^2<1,4,9,\ldots36$.}
\end{figure}

\section{Physical Parameters\label{sec:phys}}

Because $\rho$ (or equivalently $t_*=\rho t_{\rm E}$) is measured, it will
be possible to determine the lens-source relative proper motion $\mu =
\theta_*/t_*$ and the angular Einstein radius,
$$\theta_{\rm E}\equiv\sqrt{\kappa M\pi_{\rm rel}},\qquad
\kappa\equiv\frac{4{\rm G}}{c^2\,{\rm a.u.}}\simeq8.14\frac{{\rm mas}}{\msun},\eqno(2)$$ 
{\it via} $\theta_{\rm E}=\theta_*/\rho$. Here, $M$ is the lens mass and
$\pi_{\rm rel}$ is the lens-source relative parallax. Unfortunately, it is
not possible to measure the microlens parallax $\pi_{\rm E}$ for this event,
which would have permitted direct determination of the mass $M=\theta_{\rm
E}/\kappa\pi_{\rm E}$ and $\pi_{\rm rel}=\theta_{\rm E}\pi_{\rm E}$ \citep{gould92,gould00}.
Hence it will be necessary to estimate these quantities from a
Bayesian analysis.

\subsection{Measurement of $\theta_*$\label{sec:thetastar}}

To determine $\theta_*$ we first measure the offset of the source star from
the Red Clump on an instrumental color--magnitude diagram.  We find the
color from regression of {\it V} on {\it I} flux based on the KMTC BLG41
data, which fortuitously has a point near the peak of the caustic exit. We
find $\Delta(V-I)=-0.300\pm0.055$~mag for the color offset from the
clump. Next, we find the offset in {\it I}-band magnitude by comparing the
source flux in the instrumental system to that of the clump. Because there
are two solutions that differ by $-2.5\log(f_{\rm s,A},f_{\rm s,B})=I_{\rm
s,A}-I_{\rm s,B}=-0.174$~mag, the offset from the clump will likewise differ
by this amount.  We find $\Delta I=+3.43\pm0.08$~mag (A) and $\Delta
I=+3.60\pm0.08$~mag (B). That is, the source is likely to be a bulge turn-off
star in either case. We adopt $[(V-I),I]_{\rm clump,0}=(1.06,14.54)$~mag
\citep{bensby13,nataf13}, to obtain $(V-I)_{s,0}=0.76\pm0.06$~mag
and $I_{s,0}=17.97\pm0.09$~mag (A) and $I_{s,0}=18.14\pm0.09$~mag (B). Then
converting from $(V-I)$ to $(V-K)$ using the color--color relations of
\citet{bb88}) and applying the color/surface-brightness
relations of \citet{kervella04}, we obtain
$$\theta_*=0.845\pm0.079~\mu {\rm mas}\quad {\rm (A)},\qquad
  \theta_*=0.781\pm0.073~\mu {\rm mas}\quad {\rm (B)}.\eqno(3)$$

\subsection{Determination of $\theta_{\rm E}$ and $\mu$\label{sec:thetae}}

Applying Eq.(3) and the results in Table~2, we then obtain
$$\theta_{\rm E}=\frac{\theta_*}{\rho}=0.288\pm0.029~{\rm mas}\quad {\rm (A)},\qquad
  \theta_{\rm E}=0.317\pm0.035~{\rm mas}\quad {\rm (B)}.\eqno(4)$$
and
$$\mu=\frac{\theta_*}{t_*}=8.81\pm 0.097~{\rm mas/yr}\quad {\rm (A)},\qquad
\mu=8.91\pm0.098~{\rm mas/yr}\quad {\rm (B)}.\eqno(5)$$

The relatively high proper motion is suggestive of a disk lens. If both the
lens and the source were in the bulge, this would imply a relative velocity
$\Delta v\simeq\mu D_S\simeq 350$~km/s, where we have adopted
$D_s=8.55$~kpc for the source distance. This would be high, but certainly
not impossible, given the one-dimensional dispersion of bulge lenses
$\sigma_{\rm bulge}\approx2.7$~mas/yr derived from OGLE-IV measurements. On
the other hand, the proper motions in Eq.(5) would be quite typical of disk
lenses.

However, the projected density of bulge stars is much higher than that of
disk stars, which can partially compensate for the relatively poor match to
bulge kinematics.

The Einstein radius places a joint constraint on the mass and relative
parallax:
$$M=\frac{\theta_{\rm E}^2}{\kappa\pi_{\rm rel}}\rightarrow 
0.11~\msun\biggl(\frac{\pi_{\rm rel}}{0.10~{\rm mas}}\biggr)^{-1}\eqno(6)$$
where we have made the evaluation using $\theta_{\rm E}\rightarrow
0.3$~mas as a representative value, and where we have normalized to a
``typical'' relative parallax for a disk lens. Thus, because the proper
motion provides only very weak constraints on the lens distance, we
expect a very broad distribution of possible masses, centered on low-mass
M dwarfs.

\subsection{Bayesian Analysis\label{sec:bayes}}

To make a more quantitative evaluation, we carry out a Bayesian analysis
with the aid of a \citet{han95} Galactic model. We draw sources and
lenses randomly from the velocity and physical distributions of that model.
For lenses, we draw according to $\nu(D_L)D_L^2dD_L$ where $D_L$ is the
lens distance, but for sources we use $\nu(D_S)dD_S$ to account for the
lower frequency of more luminous sources (required to generate the observed
source flux at greater distance). However, we find that this correction
has a very small impact on the posterior distributions.  Our results are
given in Table~3 for each of the two solutions. We note that the host mass
and distance estimates are statistically indistinguishable between the two
models, but the planet mass is a factor 2.2 higher in Local A, simply
because $q$ is higher by a similar factor.

\begin{deluxetable}{lcc}
\tablecolumns{3} 
\tablewidth{0pc}
\tablecaption{\textsc{Physical properties}} 
\tablehead{\colhead{Quantity}&\colhead{Local A}&\colhead{Local B}}
\startdata
  $M_{\rm host}$ $[M_\sun]$    &$0.248_{-0.133}^{+0.274}$  &$0.276_{-0.148}^{+0.296}$ \\
  $M_{\rm planet}$ $[M_J]$     &$0.401_{-0.257}^{+0.637}$  &$0.189_{-0.116}^{+0.270}$\\
  $D_{\rm L}$ [kpc]           &$5.915_{-1.478}^{+1.123}$  &$5.810_{-1.523}^{+1.133}$ \\
  $a_\bot$ [au]              &$2.424_{-0.783}^{+0.757}$  &$2.575_{-0.878}^{+0.850}$ \
\enddata
\label{tab:phys}
\end{deluxetable}

Hence, as anticipated in Section~4.2, a quite broad range of host and
planet masses are consistent with the microlensing solutions.

\begin{figure}
\plotone{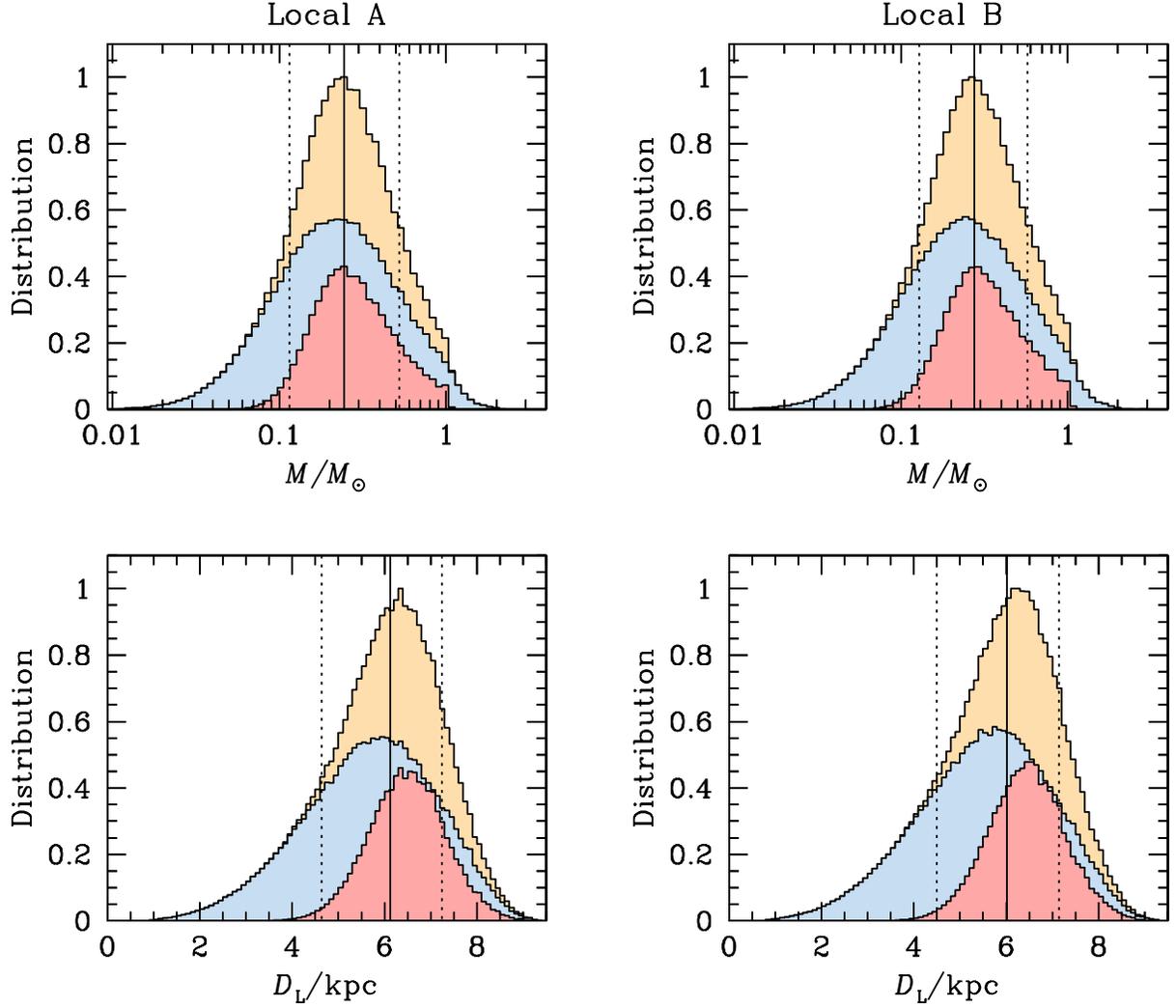}
\caption{Posterior distributions of the host mass and system distance of
  OGLE-2017-BLG-0373Lb, which prove to be nearly identical for the two
  degenerate solutions, A and B. The total probability distribution (gold)
  is divided into disk (blue) and bulge (red) lenses.  The median (50\%),
  16\%, and 84\% percentiles are shown.  The true values of these host
  properties can easily be determined from high-resolution imaging
  $\approx10$~yr after the event. However, resolving the factor 2.5
  mass-ratio degeneracy between the two models will be more
  challenging. See text.}
\end{figure}

\section{Discussion\label{sec:discuss}}

The degeneracy presented here was relatively easy to discover because the
$\chi^2$ barrier between the two solutions is modest $\Delta\chi^2_{\rm
barrier}\approx9$ (see Fig.~5). Hence, it can in principle appear in the
course of running an MCMC to evaluate the error bars of the first solution
that is found, or alternatively in a mildly hotter MCMC. This situation is
more favorable for discovery than the case of OGLE-2017-BLG-0173, which had
a much higher barrier \citep{ob170173}.  However, this ``favorable''
circumstance is actually a by-product of unusually noisy KMTNet data for an
event of this brightness.  The source star happened to lie below two
relatively bright stars on the CCD, which bled by small amounts onto the
point spread function (PSF) of the source. If this were not the case, the
error bars would have been smaller by a factor $\approx1.7$, implying that
$\Delta\chi^2_{\rm barrier}$ would have been $\approx3$ times higher.

Thus, as in the cases outlined in Section~1, it is important to have an
analytic understanding of this degeneracy in order to recognize when it
might be present and to know how to look for it. We have provided such
guidelines in Section~3.3.

\subsection{Future Resolution\label{sec:resolve}}

Future high-resolution images can clearly resolve the uncertainties in the
host mass and distance that were evaluated in Section~4.3, but the problem
of breaking the factor $\approx2.5$ discrete degeneracy in $q$ (and so in
planet mass) is considerably more challenging.

Using Keck adaptive optics (AO), \citet{ob05169bat} fully resolved the
source and lens of OGLE-2005-BLG-169 when they were separated by
$\approx60$~mas, and \citet{ob05169ben} partially resolved them when they
were separated by $\approx50$~mas using the Hubble Space Telescope
(HST). That case was somewhat more favorable than the present one because
the source and lens had comparable fluxes, whereas for OGLE-2017-BLG-0373
the lens could be substantially fainter than the source.  Therefore, to be
conservative, we can estimate that full resolution with presently existing
instruments should be possible when the separation is
$\approx90$~mas. Given that the proper motion is $\mu\approx9$~mas/yr, such
imaging could be done 10 years after the event. Combining such imaging with
the mass-distance relation for the host derived from the $\theta_{\rm E}$
measurement (Eq.~6) would lead to a precise host-mass measurement as it did
for OGLE-2005-BLG-169.

There is also the potential for future high-resolution imaging to resolve
the discrete degeneracy between solutions A and B. From Table~2, these two
solutions predict {\it I}-band magnitudes that differ by $I_{\rm
  s,A}-I_{\rm s,B}=-0.174\pm0.067$~mag. This is a $2.6\sigma$ difference,
so in principle if the source flux were accurately measured in future
high-resolution images, this could resolve the mass-ratio degeneracy with
reasonable confidence\footnote{In order to enable such an excercise,
  photometry of neighbouring stars (in the same photometric system as the
  light curve) will be submitted to NASA Exoplanet Archive and available
  from {\it https://exoplanetarchive.ipac.caltech.edu/} as well as from the
  OGLE website ({\it http://ogle.astrouw.edu.pl/}).}. Unfortunately, our
measurement of the source color (Section~4.1) was relatively imprecise,
which implies that to determine the {\it I}-band magnitude, one must make
the measurement in a bandpass near {\it I}-band and also measure the source
flux in a second relatively nearby bandpass to determine the color. Then
one would be able to accurately estimate $I_s$ using color--color relations
based on field stars.  Hence, this measurement probably requires either
HST, which will probably be decommissioned by that time, or the James Webb
Space Telescope, assuming that the latter is successfully launched and
remains operational.

\acknowledgments 
  The OGLE project has received funding from the National Science Centre,
  Poland, grant MAESTRO 2014/14/A/ST9/00121 to AU.

  Work by WZ, YKJ, and AG were supported by AST-1516842 from the US NSF.
  WZ, IGS, and AG were supported by JPL grant 1500811.

  This research has made use of the KMTNet system operated by the Korea
  Astronomy and Space Science Institute (KASI) and the data were obtained
  at three host sites of CTIO in Chile, SAAO in South Africa, and SSO in
  Australia.

  Work by YS was supported by an appointment to the NASA Postdoctoral
  Program at the Jet Propulsion Laboratory, California Institute of
  Technology, administered by Universities Space Research Association
  through a contract with NASA.

  Work by C. Han was supported by grant (2017R1A4A1015178) of the National
  Research Foundation of Korea.

\end{document}